\documentclass{article}
\usepackage[top=1in, bottom=1in, left=1in, right=1in]{geometry}
\usepackage{latexsym}
\usepackage{graphicx}
\usepackage{mathptmx}
\usepackage[pdftex,colorlinks=true,urlcolor=blue,citecolor=black,anchorcolor=black,linkcolor=black]{hyperref}
\usepackage{amsmath}

\usepackage{listings}
\usepackage{color}
\usepackage[subdued,defaultmathsizes]{mathastext}

\MTnonlettersobeymathxx     
\MTexplicitbracesobeymathxx 
\MTfamily {\ttdefault}      
\Mathastext [typewriter]    
\definecolor{dkgreen}{rgb}{0,0.6,0}
\definecolor{gray}{rgb}{0.5,0.5,0.5}
\definecolor{mauve}{rgb}{0.58,0,0.82}
\newcommand{\closeup}{\setlength{\itemsep}{-4pt}}
\def\calF/{${\cal F}$}
\def\calR/{${\cal R}$}
\def\pram/{\textsc{pram}}
\def\adams/{\texttt{Adams}}
\def\berry/{\texttt{Berry}}
\def\home/{\texttt{Home}}

\def\g/{\texttt{g}}

\newenvironment{inl}[1]
    {$\mathtt{#1}$}

\newenvironment{grp}[1]
    {$\mathtt{g_{#1}}$}

\begin{document}

\title{Redistribution Systems and PRAM}

\author{
\\
Paul Cohen \\
Tomek Loboda \\ [12pt]
School of Computing and Information \\
University of Pittsburgh \\
Pittsburgh, PA, USA \\
prcohen@pitt.edu}

\maketitle
\abstract{ 
\noindent Redistribution systems iteratively redistribute mass between groups under the control of rules.  \pram/ is a framework for building redistribution systems.  We discuss the relationships between redistribution systems, agent-based systems, compartmental models and Bayesian models. \pram/ puts agent-based models on a sound probabilistic footing by reformulating them as redistribution systems.  This provides a basis for integrating agent-based and probabilistic models.  \pram/ extends the themes of probabilistic relational models and lifted inference to incorporate dynamical models and simulation. We illustrate \pram/ with an epidemiological example. }

\section{Introduction}

Every Autumn, a new freshman class enters our university.  Some drop out during the year, but most go on to become sophomores. Some get their general education requirements out of the way, others jump into their major areas of study. By the end of their first year, the incoming class is distributed among several groups.  The dynamics of this distribution, month by month, year by year, depends on many factors and can be hard to analyze.  Some students take a few classes in a major and decide that it isn't what they'd hoped for.  We can model this straightforwardly as a conditional probability of sticking with the major given one's experiences in it. But if the major has limited capacity, then the number of students who stick with a major affects the number students who enter it. This is more difficult to model because the probabilities of transitions in and out of a major change over time. At our university and others there is an ongoing {\em redistribution} of students among academic, social, and other groups. New groups emerge: computer science students who are supported by the GI Bill, nursing students with minors in information science, and so on.  

\pram/ is a framework for building redistribution models and simulating their dynamics. In \pram/ models, groups are defined by attributes and the dynamics of redistribution are generated by rules that probabilistically change attribute values.  \pram/ modelers specify these rules and some initial groups, but they need not anticipate all possible groups; \pram/ generates groups automatically.  \pram/ grows and shrinks groups by redistributing their masses to other groups, some of which emerge during a \pram/ simulation.  

We built \pram/ to unify several kinds of models in a single framework.  \pram/ incorporates aspects of compartmental models (e.g.,~\cite{Blackwood2018}), agent-based models (ABMs, e.g.,~\cite{Kravari2015,Grefenstette2013}) and probabilistic relational models (PRMs; e.g.,~\cite{Getoor2007}).  Simulation of \pram/ models is a kind of lifted inference~\cite{}. We suspect that all these kinds of models are fundamentally very similar~\cite{}. \pram/ seeks to clarify the probabilistic inference done by agent-based simulations as a first step toward integrating probabilistic and agent-based methods, enabling new capabilities such as automatic compilation of probabilistic models from simulation specifications, replacing or approximating expensive simulations with inexpensive probabilistic inference, and unifying ABMs with important methods such as causal inference.  

\section{An Example}

Consider the spread of influenza in a population of students at two synthetic schools, \adams/ and \berry/. To simplify the example, assume that flu spreads only at school.  Many students at \adams/ have parental care during the day, so when they get sick they tend to recover at home.  Most students at \berry/  lack parental care during the day, so sick students go to school.  Students may be susceptible, exposed or recovered.  

\begin{figure}[hbt]
\begin{center}
\includegraphics[width=6in]{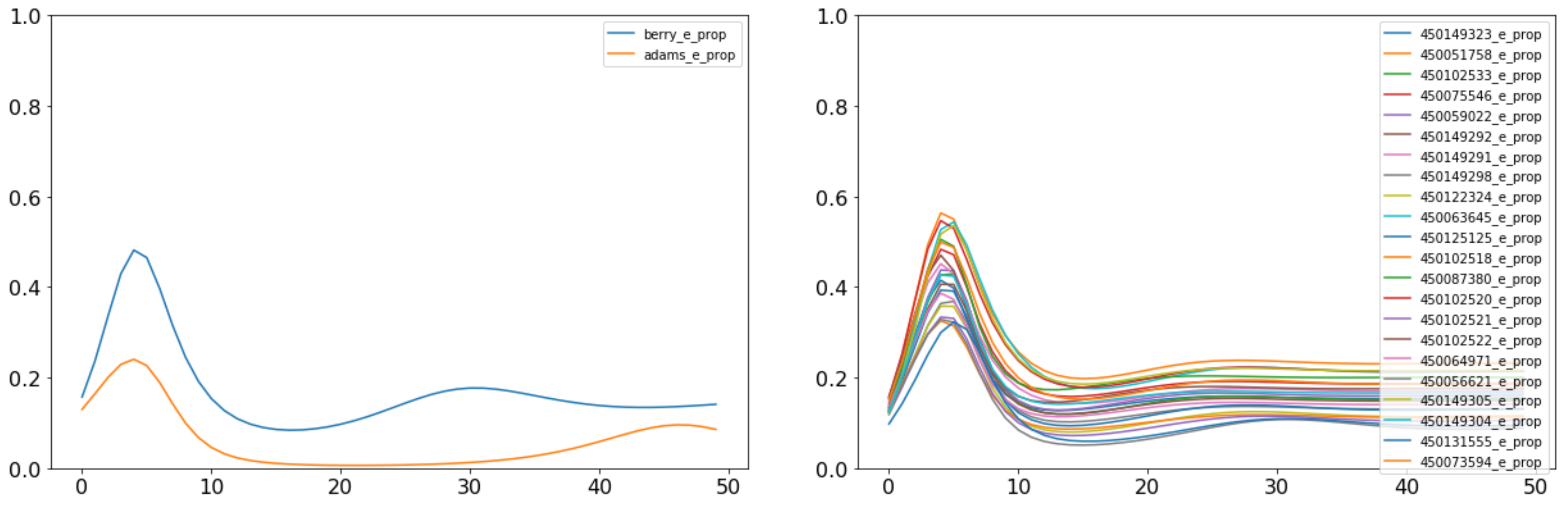}
\caption{The left panel shows the proportions of students exposed to flu at the artificial \adams/ and \berry/ schools over 50 time steps.  The right panel shows 23 the proportions of exposed students at 23 schools in Pittsburgh. 
}
\label{fig:adamsberry}
\end{center}
\end{figure} 

Although \adams/ and \berry/ are identical in all respects other than the availability of parental care, the dynamics of flu, as simulated by \pram/,  are different at the schools.  This is shown in the left side of Figure~\ref{fig:adamsberry}.  The reasons are that the probability of contracting flu at school depends on proportion of people who have it, and 80\% of \berry/ students go to school when they are sick, while 60\% of \adams/ students stay home.  
Similar, dynamics are seen for 23 schools in Pittsburgh.  In this case, we specified that the probability of going home when sick is 0.9 for a pre-schooler, 0.5 for a middle-schooler and .1 for a high-school student.

\pram/ redistributes the student populations in these examples between several groups.  There are susceptible, exposed and recovered groups; and these levels of \inl{flu\_status} are crossed with location -- home or school -- and also with particular schools -- \adams/ or \berry/ in the first example and 23 schools in the second. Indeed, in the second example, \pram/ begins with 433 groups and generates 2064 more groups as it simulates the dynamics of flu within schools.     

\section{Elements of PRAM Models}

\pram/ models comprise entities and rules.  At present, entities are {\em groups} or {\em sites}. Groups  have {\em counts} that are redistributed among groups, and they have two kinds of attributes:  unary {\em features}, \calF/, such as \inl{flu\_status} and \inl{sex}, and  binary {\em relations}, \calR/ such as \inl{has\_location}.   Groups are related to sites and sites {\em aggregate} information about the groups to which they are related in the sense that the term is used in ~\cite{}.  For example, a site might calculate the total mass of related groups that are exposed to flu.  All {\em forward} relations between groups and sites, such as \inl{g_1.has\_school = Adams} relate one group to one site. Inverse relations relate one site to a set of groups.  Thus, if \inl{g_1.has\_school = Adams} and \inl{g_2.has\_school = Adams}, the inverse relation \inl{Adams.school\_of} returns \inl{\{g_1,g_2\}}.  Inverse relations are important for answering queries such as ``which groups attend \grp{1}'s school?"  Formally this would be \inl{g_1.has\_school.school\_of}, which would return \inl{\{g_1,g_2\}}.  By mapping over entities it is easy to answer queries such as ``what is the proportion of students at \grp{1}'s school that has been exposed to flu?"  In effect, \pram/ implements a simple relational database.

Besides entities, \pram/ models have rules that apply to groups.  All rules have mutually exclusive conditions, and each condition is associated with a probability distribution over mutually exclusive and exhaustive conjunctive actions. Thus, a rule will return exactly one distribution of conjunctive actions or nothing at all if no condition is true. For an illustration, look at the mutually exclusive clauses of  \inl{rule\_flu\_progression} in Figure~\ref{rules}, and particularly at the middle clause:  It tests whether the group's \inl{flu\_status == e} (exposed to flu) and it specifies a distribution over three conjunctive actions.  The first, which has probability $0.2$, is that the group recovers {\em and} becomes happy (i.e., change \inl{flu\_status} to {\tt r} {\em and} change \inl{mood} to {\tt happy}).  The remaining probability mass is divided between remaining exposed and becoming bored, with probability $0.5$, and remaining exposed and becoming annoyed, with probability $0.3$.

\begin{figure}
\begin{center}
\begin{lstlisting}
def rule_flu_progression (group):
    flu_status = group.get_feature('flu')
    location = objects_related_by(group,'has_location')
    infection_probability = location.proportion_located_here([('flu','e')])

    if flu_status == 's':
        return ((infection_probability,
                 ('change_feature','flu','e'),('change_feature','mood','annoyed')),
                ((1 - infection_probability),('change_feature','flu','s')))  
    elif flu_status == 'e':
        return ((.2, ('change_feature','flu','r'),('change_feature','mood','happy')),
                 (.5, ('change_feature','flu','e'),('change_feature','mood','bored')), 
                 (.3, ('change_feature','flu','e'),('change_feature','mood','annoyed')))   
    else flu_status == 'r':
        return ((.9, ('change_feature','flu','r')),
                 (.1, ('change_feature','flu','s')) 
                 
def rule_flu_location (group):
    ...        
    if flu_status == 'e' and income == 'l':
        return ((.1, ('change_relation','has_location',location,home)),
                 (.9, ('change_relation','has_location',location,location)))    
    elif flu_status == 'e' and income == 'm':
        return ((.6, ('change_relation','has_location',location,home)),
                 (.4, ('change_relation','has_location',location,location)))              
    else flu_status == 'r':
        return ((.8, ('change_relation','has_location',location,school)),
                 (.2, ('change_relation','has_location',location,location)))

\end{lstlisting}
\caption{Two \pram/ rules.  \inl{Rule\_flu\_progression} changes the \inl{flu\_status} and \inl{mood} features of a group. \inl{Rule\_flu\_location} changes a group's \inl{has\_location} relation.}
\label{rules}
\end{center}
\end{figure}

Next, consider the preamble of \inl{rule\_flu\_progression}, which queries the group's flu status, then finds the group's location, and then calls the method \inl{proportion\_located\_here} to calculate the proportion of flu cases at the location. (\inl{Proportion\_located\_here} sums the counts of groups at the location that have flu, then divides by the sum of the counts of all the groups at the location.)  In the rule's first clause, this proportion serves as a probability of infection. It is evaluated anew whenever the rule is applied to a group.  In this way, rules can test conditions that change over time. 
Finally, the third clause of the rule represents the  transition from \inl{flu\_status = r} back to \inl{flu\_status = s}, whereupon re-exposure becomes possible. 

In addition to changing groups' features, rules can also change relations such as {\tt has\_location}. The second rule in Figure~\ref{rules} says, if a group is exposed to flu and is {\em low}-income then change the group's location from its current \inl{location} to \inl{home} with probability $0.1$ and stay at \inl{location} with probability $0.9$.  If, however, the group is exposed and is {\em middle}-income, then it will go home with probability $0.6$ and stay put with probability $0.4$.  And if the group has recovered from flu, whatever its income level, then it will go back to school with probability $0.8$.

\section{Groups are defined by their attributes}
\label{sec:engine}

\pram/ groups are defined by their features and relations in the following sense:  Let \calF/ and \calR/ be features and relations of group {\tt g}, and let $n$ be the count of {\tt g}.  For groups \grp{i} and \grp{j}, if \calF/$_i =$ \calF/$_j$ and \calR/$_i =$ \calR/$_j$, then \pram/ will merge \grp{i} with \grp{j} and give the result a count of $n_i + n_j$.  Conversely, if a rule specifies a distribution of $k$ changes to \calF/$_i$ (or \calR/$_i$) that have probabilities $p_1,p_2,...,p_k$, then \pram/ will create $k$ new groups with the specified changes to \calF/$_i$ (or \calR/$_i$) and give them counts equal to $(p_1 \cdot n_i), (p_2 \cdot n_i), ..., (p_k \cdot n_i)$. 

\begin{figure}[hbt]
\begin{center}
\includegraphics[width=1.5in]{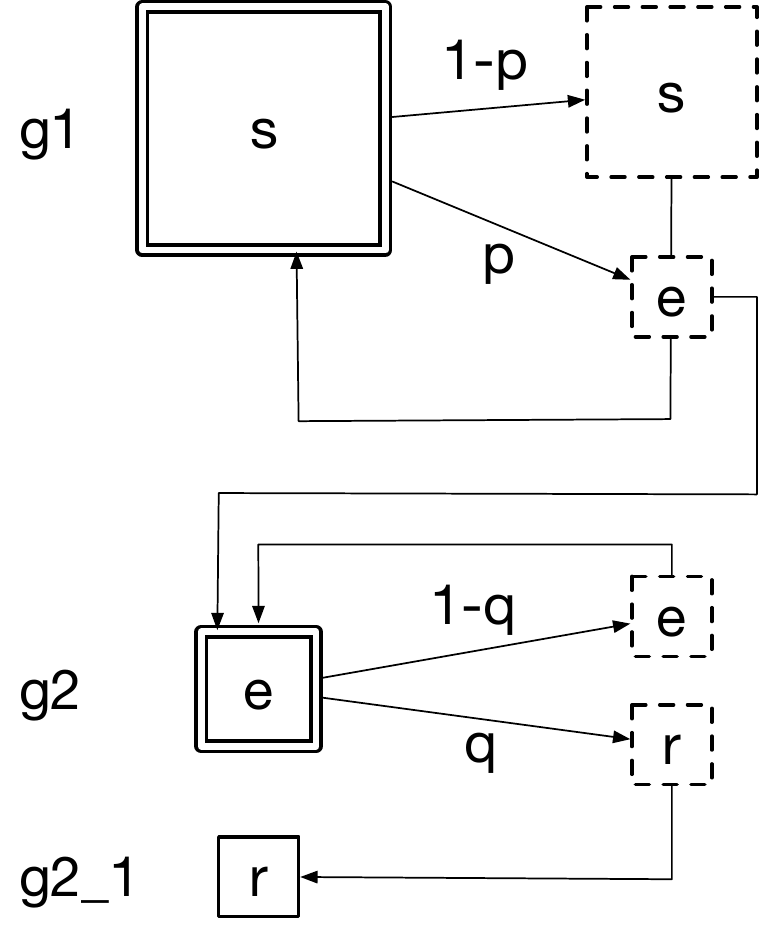}
\caption{How \pram/ splits, merges and creates groups to redistribute group counts.}
\label{fig:flow}
\end{center}
\end{figure}

To illustrate, consider a \pram/ system with just a single attribute, \inl{flu\_status}, which takes values s, e and r.  Figure~\ref{fig:flow} illustrates how groups are created, split and merged, and how their counts change.  
\vspace{.1in}

Suppose \pram/ starts with two groups \grp{1} and \grp{2} (denoted by double-lined boxes) with \inl{flu\_status = s} and \inl{flu\_status = e}, and counts $n_1$ and $n_2$, respectively.  A rule specifies that susceptible people become exposed with probability $p$, so \pram/ generates two potential groups (denoted by dotted lines) and redistributes the count of \grp{1} between them in proportions $p,1-p$.  As groups in this simple example are defined by a single feature, these potential groups are identical with \grp{2} and \grp{1}, respectively, so \pram/ will redistribute $n_1$ to \grp{2} and \grp{1}. Redistribution means that the {\em entire} count of a group, $n_1$ in this case, is distributed, so \grp{1}'s new count will be $n_1 \cdot (1 - p)$ while the count of \grp{2} will be incremented by $n_1 \cdot p$.  However, something similar is going on with \grp{2}:  A rule specifies that some exposed people will recover with probability $q$, so \pram/ spawns two potential groups with counts of $n_2 \cdot q$ and $n_2 \cdot (1 - q)$, and distributes the first to the new recovered group and the second back to \grp{2}. Finally, because the potential group labeled {\tt r} doesn't already exist, \pram/ makes it a real group (with a solid line) and gives it the name \grp{2\_1}, denoting that it is the first real group created by the action of rules on group \grp{2}.  After all this, the counts for the groups are:  
\vspace{-.3in}

\begin{align*}\closeup
\mathtt{g_1}:&~~ n_1 \cdot (1 - p)\\ 
\mathtt{g_2}:&~~ n_1 \cdot p + n_2 \cdot (1 - q)\\ 
\mathtt{g_{2\_1}}:&~~ n_2 \cdot q
\end{align*}

Clearly, the system in Figure~\ref{fig:flow} can be iterated with these counts as a new starting point. Repeated iterations will yield the dynamics of group counts. 

\pram/ isn't necessary for this simple example, which mirrors the $SIR$ compartmental developed by Kermack and McKendrick in 1927 and is well understood~\cite{Kermack1927}.  However, \pram/ handles vastly more complicated models, allowing more features, more groups, relations between groups, multiple rules applying simultaneously to groups, and nonstationary probabilities. \pram/ guarantees that group counts always obey the probabilities associated with rules, and that the order of rules and clauses within rules, and the order of application of rules to groups, have no effects on counts.  
 
\section{The PRAM Engine:  Redistributing Group Counts}
\label{sec:engine}
    
The primary function of the \pram/ engine is to redistribute group counts among groups, as directed by rules, merging and creating groups as needed, in a probabilistically sound way. To illustrate the details of how \pram/ redistributes counts, suppose a \pram/ model starts with just the two rules in Figure~\ref{rules} and two {\em extant} groups:

\begin{center}
\texttt{
\begin{tabular}{lcccr}
name & flu & mood & location & count \\ \hline
\grp{1} & s & happy & adams & 900 \\
\grp{2} & e & annoyed & adams & 100 
\end{tabular}
}\end{center}
\vspace{.075in}

The features for these groups are ${\cal F}_1$ = \inl{[flu = s, mood = happy]} and ${\cal F}_2$ = \inl{[flu = e, mood = annoyed]}, and both groups have the same relation:  ${\cal R}_1 = {\cal R}_2 =$\inl{[has\_school~adams]}.

\paragraph{Redistribution Step 1: Generate Potential Groups} When {\tt rule\_flu\_progression} is applied to \grp{1} it calculates the \inl{infection\_probability} at \inl{adams} to be $100 / (100 + 900) = .1$.  \grp{1} triggers the first clause in the rule because \grp{1}'s \inl{flu\_status == s}. So the rule specifies that the \inl{flu\_status} of \grp{1} changes to {\tt e} with probability $0.1$ and changes to {\tt s} with probability $0.9$.  \pram/ then creates two {\em potential groups}:

\begin{center}
\texttt{
\begin{tabular}{lcccr}
name & flu & mood & location & count \\ \hline
\grp{1\_1} & e & annoyed & adams & 90 \\
\grp{1\_2} & s & happy & adams & 810 
\end{tabular}
}\end{center}
\vspace{.075in}

These potential groups specify a {\em redistribution} of $n_1$, the count of \grp{1}. We will see how \pram/ processes redistributions, shortly. 

Of the two rules described earlier, {\tt rule\_flu\_location} does not apply to \grp{1}, but both apply to group \grp{2}. When multiple rules apply to a  group, \pram/ creates the cartesian product of their distributions of actions and multiplies the associated probabilities accordingly, thereby enforcing the principle that rules' effects are independent. (If one wants dependent effects they should be specified {\em within} rules.) To illustrate, {\tt rule\_flu\_progression} specifies a distribution of three  actions for groups like \grp{2} that have \inl{flu\_status=e}, with associated probabilities $0.2,0.5,0.3$; while {\tt rule\_flu\_location} specifies two locations for groups that have \inl{flu\_status=e} and \inl{flu\_status=m}, with probabilities $0.6$ and $0.4$.  Thus, for \grp{2}, there are six joint actions of these two rules, thus six potential groups: 

\begin{center}
\texttt{
\begin{tabular}{lcccl}
name & flu & mood & location & count \\ \hline
\grp{2\_1} & r & happy & home & 100 $\cdot$ 0.2 $\cdot$ 0.6 = 12.0 \\
\grp{2\_2} & r & happy & adams & 100 $\cdot$ 0.2 $\cdot$ 0.4 = 8.0 \\
\grp{2\_3} & e & bored & home & 100 $\cdot$ 0.5 $\cdot$ 0.6 = 30.0 \\
\grp{2\_4} & e & bored & adams & 100 $\cdot$ 0.5 $\cdot$ 0.4 = 20.0 \\
\grp{2\_5} & e & annoyed & home & 100 $\cdot$ 0.3 $\cdot$ 0.6 = 18.0 \\
\grp{2\_6} & e & annoyed & adams & 100 $\cdot$ 0.3 $\cdot$ 0.4 = 12.0 \\
\end{tabular}
}\end{center}
\vspace{.075in}

These groups redistribute the count of \grp{2} (which is 100) by multiplying it by the product of probabilities associated with each action. 

\paragraph{Redistribution Step 2: Process Potential Groups} \pram/  applies all rules to all groups, collecting potential groups as it goes along.  Only then does it redistribute counts, as follows:  
\begin{enumerate}
\item Extant groups that spawn potential groups have their counts set to zero;
\item Potential groups that match extant groups (i.e., have identical \calF/s and \calR/s) contribute their counts to the extant groups and are discarded;
\item Potential groups that don't match extant groups become extant groups with their given counts.
\end{enumerate} 

So: Extant groups \grp{1} and \grp{2} have their counts set to zero.  Potential group \grp{1\_2} has the same features and relations as \grp{1} so it contributes its count, 810, to \grp{1} and is discarded.  Likewise, potential group \grp{1\_1} matches \grp{2} so it contributes 90 to \grp{2} and is discarded. Potential group \grp{2\_6} also matches \grp{2}, so it contributes 12 to \grp{2} and is discarded, bringing \grp{2}'s total to 102. Potential groups \grp{2\_1}, \grp{2\_2}, \grp{2\_3}, \grp{2\_4}, and \grp{2\_5} do not match any extant group, so they become extant groups.  The final redistribution of extant groups \grp{1} and \grp{2} is:

\begin{center}
\texttt{
\begin{tabular}{lcccr}
name & flu & mood & location & count \\ \hline
\grp{1} & s & happy & adams & 810.0 \\
\grp{2} & e & annoyed & adams & 102.0 \\
\grp{2\_1} & r & happy & home & 12.0 \\
\grp{2\_2} & r & happy & adams & 8.0 \\
\grp{2\_3} & e & bored & home & 30.0 \\
\grp{2\_4} & e & bored & adams & 20.0 \\
\grp{2\_5} & e & annoyed & home & 18.0 \\
\end{tabular}
}\end{center}

\vspace{.075in}

By delaying the processing of potential groups until all rules have been applied to all extant groups, \pram/ avoids order effects.  Imagine that \pram/ applied {\tt rule\_flu\_progression} to \grp{1} and immediately processed the resulting potential groups, and then applied the rule to \grp{2}.  Processing potential group \grp{1\_1} would make $n_2 = 100 + 90$, and applying the rule to \grp{2} would redistribute 190 between \grp{2\_1} and \grp{2\_2}.  Whereas, processing the groups in the opposite order would redistribute 80 between \grp{2\_1} and \grp{2\_2}. \pram/ eliminates effects of the order of processing of groups.  It also eliminates effects of the order of application of rules to groups, as we shall see. 

\paragraph{Redistribution Step 3: Iterate}  \pram/ is designed to explore the dynamics of group counts, so it generally will run iteratively.  At the end of each iteration, all non-discarded groups are marked as extant and the preceding steps are repeated: All rules are applied to all extant groups, all potential groups are collected, potential groups that match extant groups are merged with them, and new extant groups are created.  A second iteration produces one such new group when the third clause of {\tt rule\_flu\_progression} is applied to \grp{2\_1}:

\begin{center}
\texttt{
\begin{tabular}{lcccr}
name & flu & mood & location & count \\ \hline
\grp{2\_1\_1} & s & happy & home & 0.24 \\
\end{tabular}
}\end{center}

\vspace{.075in}

The reader is invited to calculate the full redistribution resulting from a second iteration (it is surprisingly difficult to do by hand).\footnote{The second iteration produces $n_1 = 706.632,~~n_2 = 119.768, ~~n_{2\_1} = 26.4, ~~n_{2\_1\_1} = 0.24, ~~n_{2\_2} = 25.6, ~~n_{2\_3} = 60.6, ~~n_{2\_4} = 24.4, ~~n_{2\_5} = 36.36$.}

\section{Discussion}

\pram/ incorporates elements of compartmental models, agent-based models, dynamic Bayesian models, Markov chain models and probabilistic relational models in a single framework.  From compartmental models it takes the idea of homogenous groups (e.g., the group of all individuals exposed to flu), but unlike in compartmental models, \pram/ allows for thousands of groups, relationships between groups, and non-stationary probabilities of transitions between groups.  Also, \pram/ generates groups automatically, whereas compartmental models require the modeler to specify all the compartments at the outset.

By working with groups rather than individuals, \pram/ implements lifted inference. The connection between \pram/ models and probabilistic models is that counts are proportional to posterior probabilities conditioned on attributes such as \inl{has\_school} and \inl{flu\_status} and on the actions of rules that change attributes. \pram/ applies rules repeatedly to groups, creating novel groups and merging identical groups, thereby simulating the dynamics of groups' counts.  It is in many respects like Probabilistic Relational Models (PRMs) in which groups are defined by their features and relations, but it is designed for simulation and for exploring the dynamics of group counts.  Updates that are handled by conditional probability tables in PRMs are handled by rules in \pram/, and the probabilities in these rules can change dynamically. Nevertheless, there are strong affinities between \pram/ models, dynamic Bayesian networks and PRMs, and we are currently working on methods to translate one into another.   

\pram/ models may also be viewed as a kind of agent-based model (ABMs) in which identical agents constitute groups. This idea offends the tenet of ABMs that agents are unique, but as a practical matter, agents are {\em not} unique.  Consider the roughly two million K12 students in Allegheny County. After mapping age to grade level, mapping nine race classes to four, mapping household size to just three levels, mapping individual households to 350 regions, and ignoring sex, we obtained just 3729 groups. For the purposes of simulating flu dynamics, these mappings are more than generous:  Flu affects girls and boys the same, so sex is irrelevant; and race classes might or might not affect flu transmission. Instead of assuming that agents are unique, \pram/ models assume that all members of a group are {\em functionally identical}. Two entities $i$ and $j$ are functionally identical if \calF/$_i =$ \calF/$_j$ and \calR/$_i =$ \calR/$_j$ after removing all features from \calF/$_i$ and \calF/$_j$ and all relations from \calR/$_i$ and \calR/$_j$ that are not mentioned in any rule.  Said differently, even if two entities have different features and relations, if these don't affect the behavior of a set of rules, then the rules treat these entities in exactly the same way.  
  
Another reason to prefer \pram/ over conventional ABMs is that the probabilistic foundations and guarantees of ABMs are murky, at best. In agent-based models agents probabilistically change state.  State can be represented as attribute values such as health status, monthly income, age, political orientation, location and so on. A population of agents has a {\em joint state} that is typically a joint distribution; for example, a population has a joint distribution over income levels and political beliefs.  ABMs are a popular method for exploring the {\em dynamics} of joint states, which can be hard to estimate when attribute values depend on each other, and populations are heterogeneous in the sense that not everyone has the same distribution of attribute values, and the principal mechanism for changing attribute values is interactions between agents. ABMs are no doubt engines of probabilistic inference, but it is difficult to say anything about the models that underlie the inference.  \pram/ seeks to clarify the probabilistic inference done by agent-based simulations.

\section{Future Work}

\pram/ code is available on github ~\cite{Loboda2019}.  It has run on much larger problems, including a simulation of daily activities in Allegheny County that involved more than 200,000 groups.  \pram/ runtimes are proportional to $\nu$ the number of groups, not the group counts, so \pram/ can be much more efficient than agent-based simulations (ABS).  Indeed, when group counts become one, \pram/ {\em is} an ABS, but in applications where agents or groups are functionally identical \pram/ is more efficient than ABS. 

Because $\nu$ depends on the numbers of features and relations, and the number of discrete values each can have, \pram/ could generate enormous numbers of groups.  In practice, the growth of $\nu$ is controlled by the number of groups in the initial population and the actions of rules.  Typically, $\nu$ grows very quickly to a constant, after which \pram/ merely redistributes counts between these groups.  In the preceding example, the initial $\nu = 8$ groups grew to $\nu = 44$ on the first iteration and $\nu = 52$ on the second, after which no new groups were added.   

This dependence between $\nu$ and the actions of rules suggests a simple idea for {\em compiling} populations given rules: Any feature or relation that is not mentioned in a rule need not be in groups' \calF/ or \calR/.  Said differently, the only attributes that need to be in groups' definitions are those that condition the actions of rules. Currently we are building a compiler for \pram/ that automatically creates an initial set of groups from two sources: A database that provides \calF/ and \calR/ for individuals and a set of rules.  The compiler eliminates from \calF/ and \calR/ those attributes that aren't queried or changed by rules, thereby collapsing a population of individuals into groups with known counts.  

\begin{figure}[htb]
\begin{center}
\includegraphics[width=4in]{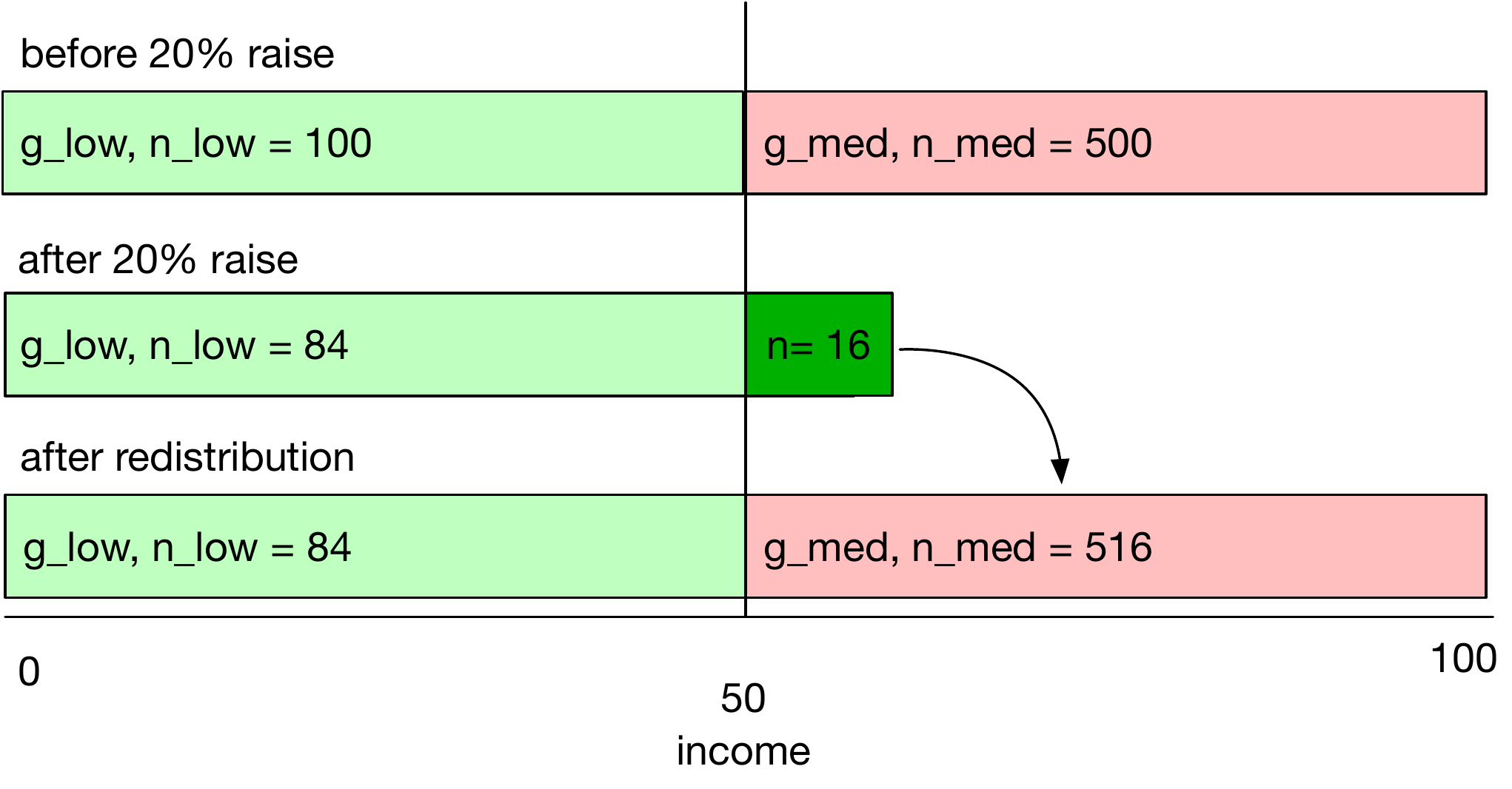}
\caption{The dividing line between low and medium income is 50.  If $g_{low}$ gets a 20\% raise, then 16\% of $g_{low}$ will make more than 50. After redistribution, this fraction of $g_{low}$ increases the count of $g_{med}$.  The count of $g_{low}$ is reduced accordingly.}
\label{fig:univariate}
\end{center}
\end{figure}

Attributes with continuous values obviously can result in essentially infinite numbers of groups. 
(Imagine one group with a single real-valued feature and one rule that adds a standard normal variate to it. Such a \pram/ model would double the number of groups on each iteration without limit.) 
Rather than ban real-valued attributes from \pram/ we are working on a method by which groups have distributions of such attributes and rules change the parameters of these distributions. We are developing efficient methods by which \pram/ generates new potential groups and tests whether they match extant groups.  To illustrate the approach, suppose we have a population distribution of income which, for the sake of simplicity, is uniform over the range [0,100].  Suppose we define two groups according to this exogenous distribution $g_{low}$ has income less than 50, whereas $g_{med}$ has income greater than or equal to 50.  Now suppose that every member of $g_{low}$ gets a 20\% percent raise.  It turns out that 16\% of $g_{low}$ will make more than 50 after the raise.  Let's call this fraction the upwardly mobile, or $UM$. Suppose $g_{low}$ has count $n_{low} = 100$ and $g_{med}$ has $n_{med} = 500$.  If income is truly uniformly distributed in $g_{low}$, then after a uniform 20\% raise, 16\% of $g_{low}$ will make more than 50.  If income were the only factor that defined groups, then PRAM redistribution should reduce the count of $g_{low}$ by 16 and increase the count of $g_{med}$ by 16, as illustrated in Figure~\ref{fig:univariate}.  

This example suggests that groups can be defined based on real-valued features:
\begin{enumerate}
\item An {\em exogenous} distribution of income is defined and divided into two {\em regions} which we'll call the low and medium income regions.  In general, we will define a multivariate distribution and divide it into many regions.  
\item Two groups are defined, each with an {\em endogenous} distribution of income; call these distributions $I_{low}$ and $I_{med}$. In general, groups are defined by the relationships between their (multivariate) endogenous distributions and the regions of the (multivariate) exogenous distribution.  One kind of relationship is ``contained in": $I_{low}$ and $I_{med}$ are contained in the low income and medium income regions, respectively.
\item A {\em labeling function} sweeps the endogenous distribution of a group and returns the relationships that hold between elements of the distribution and regions.  For example, before the raise, the labeling function would say that all the element of $I_{low}$ are contained in the low income region.
\item A rule would change $I_{low}$ by multiplying $I_{low}$ by 1.2.
\item The labeling function would relabel $I_{low}$.  This time, it would label 84\% of the distribution as contained in the low income region and 16\% as contained in the medium income region.  
\item PRAM would use these labels as features and so would split $g_{low}$ into two groups:  One would be re-merged with $g_{low}$, the other would merge with $g_{med}$.
\end{enumerate}

We are currently working on this mechanism for groups that are defined are regions of multivariate distributions constructed automatically from databases. 

In sum, while \pram/ is a simple algorithm for redistributing counts of groups, it appears to unify several other modeling frameworks.  The primary advantage of \pram/ over ABS is that \pram/ models are guaranteed to handle probabilities properly.  The steps described in Section~\ref{sec:engine} ensure that group counts are consistent with the probability distributions in rules and are not influenced by the order in which rules are applied to groups, or the order in which rules' conditions are evaluated. These guarantees are the first step toward a seamless unification of databases with probabilistic and \pram/ models.  The next steps, which we have already taken on a very small scale, are automatic compilation of probabilistic models given \pram/ models, and automatic compilation of \pram/ rules given probabilistic models. Probabilistic relational models, which inspired \pram/, integrate databases with lifted inference in Bayesian models; \pram/ adds simulation to this productive mashup, enabling models of dynamics.  

\section{Acknowledgments}
This work is funded by the DARPA program ``Automating Scientific Knowledge Extraction (ASKE)'' under Agreement HR00111990012 from the Army Research Office.


\begin{thebibliography}{9}
\bibitem{Blackwood2018} 
Julie C. Blackwood and Lauren M. Childs. An introduction to compartmental modeling for the budding infectious disease modeler. Letters in Biomathematics, 5(1), pp.195-221. 2018. doi:10.1080/23737867.2018.1509026

\bibitem{Getoor2007}
Lise Getoor, Ben Taskar (Eds.) Introduction to Statistical Relational Learning.  2007. MIT Press

\bibitem{Grefenstette2013}
Grefenstette JJ, Brown ST, Rosenfeld R, Depasse J, Stone NT, Cooley PC, Wheaton WD, Fyshe A, Galloway DD, Sriram A, Guclu H, Abraham T, Burke DS. FRED (A Framework for Reconstructing Epidemic Dynamics): An open-source software system for modeling infectious diseases and control strategies using census-based populations. BMC Public Health, 2013 Oct;13(1), 940. doi: 10.1186/1471-2458-13-940.

\bibitem{Kermack1927}
Kermack WO, McKendrick AG (August 1, 1927). ``A Contribution to the Mathematical Theory of Epidemics". Proceedings of the Royal Society A. 115 (772): 700?721. doi:10.1098/rspa.1927.0118

\bibitem{Kravari2015}
Kalliopi Kravari and Nick Bassiliades. A Survey of Agent Platforms.  2015.

\bibitem{Loboda2019} 
Loboda, T. The version of \pram/ reported here was developed by the first author.  A better engineered version has been developed by Tomek Loboda: \url{https://github.com/momacs/pram/} with documentation at \url{https://github.com/momacs/pram/blob/master/docs/Milestone-3-Report.pdf}. 2019
\end{thebibliography}
\end{document}